\documentclass[twocolumn]{aastex631}
\usepackage{graphicx,bm,amssymb,amsmath,times}
\numberwithin{equation}{section}
\usepackage{gensymb}

\newcommand{\Msun}{{\rm M}_{\odot}}
\def\gtsim {>\kern-1.2em\lower1.1ex\hbox{$\sim$}~}   
\def\ltsim {<\kern-1.2em\lower1.1ex\hbox{$\sim$}~}   

\received{November 8, 2022}
\revised{December 7, 2022}
\accepted{December 18, 2022}
\submitjournal{ApJL}

\shorttitle{Can neutron star mergers alone explain the r-process enrichment of the Milky Way?}
\shortauthors{Kobayashi, Mandel et al.}

\begin{document}
\title{Can neutron star mergers alone explain the r-process enrichment of the Milky Way?}

\author[0000-0002-4343-0487]{Chiaki Kobayashi}
\affiliation{Centre for Astrophysics Research, Department of Physics, Astronomy and Mathematics, University of Hertfordshire, Hatfield, AL10 9AB, UK}
\email{c.kobayashi@herts.ac.uk}

\author{Ilya Mandel}
\affiliation{School of Physics and Astronomy, Monash University, Clayton, VIC 3800, Australia}
\affiliation{The ARC Centre of Excellence for Gravitational Wave Discovery -- OzGrav}
\email{ilya.mandel@monash.edu}

\author{Krzysztof Belczynski}
\affiliation{Nicolaus Copernicus Astronomical Center, Polish Academy of Sciences, ul. Bartycka 18, 00-716 Warsaw, Poland}

\author{Stephane Goriely}
\affiliation{Institut d'Astronomie et d'Astrophysique, CP-226, Universit{\'e} Libre de Bruxelles, 1050 Brussels, Belgium}

\author{Thomas H. Janka}
\affiliation{Max-Planck-Institut f{\"u}r Astrophysik, Postfach 1317, 85741 Garching, Germany}

\author{Oliver Just}
\affiliation{GSI Helmholtzzentrum f{\"u}r Schwerionenforschung, Planckstrasse 1, 64291 Darmstadt, Germany}
\affiliation{Astrophysical Big Bang Laboratory, RIKEN Cluster for Pioneering
Research, 2-1 Hirosawa, Wako, Saitama 351-0198, Japan}

\author{Ashley J. Ruiter}
\affiliation{School of Science, University of New South Wales Canberra,
The Australian Defence Force Academy, 2600 ACT, Canberra, Australia}

\author{Dany Vanbeveren}
\affiliation{Department of Physics and Astronomy, Vrije Universiteit Brussel, Belgium}

\author{Matthias U. Kruckow}
\affiliation{D\'{e}partement d'Astronomie, Universit\'{e} de Gen\`{e}ve, Chemin Pegasi 51, CH-1290 Versoix, Switzerland}

\author{Max M. Briel}
\affiliation{Department of Physics, Faculty of Science
The University of Auckland, New Zealand}

\author{Jan J. Eldridge}
\affiliation{Department of Physics, Faculty of Science
The University of Auckland, New Zealand}

\author{Elizabeth Stanway}
\affiliation{Department of Physics, University of Warwick, Gibbet Hill Road, Coventry CV4 7AL, UK}

\begin{abstract}
Comparing Galactic chemical evolution models to the observed elemental abundances in the Milky Way, we show that neutron star mergers can be a leading r-process site only if at low metallicities such mergers have very short delay times and significant ejecta masses that are facilitated by the masses of the compact objects.
Namely, black hole-neutron star mergers, depending on the black-hole spins, can play an important role in the early chemical enrichment of the Milky Way. We also show that none of the binary population synthesis models used in this paper, i.e., COMPAS, StarTrack, Brussels, ComBinE, and BPASS, can currently reproduce the elemental abundance observations. The predictions are problematic not only for neutron star mergers, but also for Type Ia supernovae, which may point to shortcomings in binary evolution models.
\end{abstract}


\section{Introduction}

Neutron star mergers (NSMs)\footnote{NSM denotes both the mergers of double neutron stars and of a neutron star and a black hole throughout this paper.} were proposed half a century ago \citep[e.g.,][]{lat74} as a production site of the rapid neutron-capture process (r-process) elements such as europium, gold, and uranium. Recently, the gravitational-wave event GW170817 \citep{abb17a}, also observed as a kilonova \citep[AT2017gfo;][]{cou17,sma17,val17,Tanvir:2017,GW170817:MMA} and a short gamma-ray burst \citep{abb17b} with X-ray, optical and radio afterglows \citep{Mooley:2017,Lyman:2018,Troja:2017}, confirmed NSMs as an r-process production site. 
The observational properties of AT2017gfo may be explained with neutron-rich dynamical ejecta, enhancing lanthanide production, and neutron-poorer outflows from the disc (\citealt{met14,cow17,pia17,tan17,Metzger:2019kilonova,kawaguchi18}, but see \citealt{Kasen:2017}), although the exact link between the kilonova observations and NSM ejecta components is not well understood.
Strontium is directly detected in the day-1 spectra of the kilonova \citep{wat19,gillanders22}. Lanthanide production was supported by the day-2 spectra, but found to be rather small in later observational analyses \citep{Waxman:2017,dom21}.
Constraining the production of the third-peak elements such as gold and platinum is challenging.

Since then one more gravitational-wave detection of a neutron star (NS)-NS merger, GW190425, has been announced \citep{abb20a}.   In addition, two mergers of a NS with a black hole (BH) have been confidently detected, GW200105 and GW200115  \citep{GW200105}; the provenance of GW190814, which could be a NS-BH or BH-BH merger, is unclear \citep{abb20b}.
The mass ejection from these BH-NS mergers is significant only for a small fraction of events \citep[perhaps $\lesssim 10$\%,][]{Drozda:2020}.
There are simulations of BH-NS mergers \citep[e.g.,][]{jan99,Rosswog:2005} but estimating the total mass ejection from BH-NS mergers requires 3D general relativistic (GR) hydrodynamical simulations with neutrino transport (\citealt{ruffert99,siegel18}; see also \citealt{fou17} and the references therein). 
The mass of ejected material depends on the equation of state (EoS) of neutron stars, the spin of the BHs \citep{bau14,kyu15} and the magnetic field \citep{pas15,kiu15}, as well as the masses of the binary components \citep{kruger20}.
The nucleosynthesis yields, however, have been provided only for a limited number of parameters in the numerical simulations (post-processing of hydrodynamical simulations; \citealt{bau14,wan14,jus15}).  

Galactic chemical evolution (GCE) models have been challenging NSMs as the main site of r-process nucleosynthesis \citep{arg04,ces15,weh15,hay19,cot19,kob20sr,van20,mol21}.
The assumed delays between star formation and the occurrence of NSMs appear to be too long to explain the observed r-process enhancement (e.g., [Eu/Fe]) of extremely metal-poor stars in the Milky Way, while the rate of NSMs may be too low.
The inclusion of inhomogeneous enrichment through state-of-the-art hydrodynamical simulations \citep{hay19,van20,van22} in some of these models leads to robust constraints on r-process enrichment, when the simulations predict the star formation and chemical enrichment history of our Milky Way Galaxy\footnote{\citet{shen15} concluded that NSM could be a dominant source of the r-process, but their simulation did not represent the solar neighborhood as the star formation timescale was too short and the observed [$\alpha$/Fe] ratios were not well reproduced.};
inhomogeneous mixing does not help resolve discrepancies between model predictions and observations of stars.
Therefore, galaxy modelling alone is unlikely to relieve the tension between observations and NSMs as the main r-process source within the current understanding of NSMs. Alternatively, r-process enrichment associated with core-collapse supernovae has been developed: magnetorotational supernovae/hypernovae \citep[e.g.,][]{nis15,yon21,rei23} or accretion disks/collapsars (\citealt{sie19}, but see \citealt{just22}).
Meanwhile, \citet{gri22} suggested a new site in mergers of neutron stars with donor cores inside a common envelope.

The NSM rate has been estimated through binary population synthesis (BPS) models \citep[e.g.,][]{men14,men16,bel16,vig18,kru18}\footnote{Among other channels to form NSMs, dynamical formation in dense stellar environments is supposed to be not important for galactic chemical evolution \citep[e.g.,][]{martell16} and the predicted number is low \citep{MandelBroekgaarden:2021}.
NSMs arising from hierarchical triples could be more significant \citep[e.g.,][]{HamersThompson:2019}.}, whose predictions broadly match the merger rate inferred from  Galactic binary pulsars, gravitational-wave events, and short gamma-ray bursts (see \citealt{MandelBroekgaarden:2021} for a review).  BPS models provide both a number of NSMs per unit star formation rate as a function of progenitor star metallicity and a delay-time distribution (DTD) between star formation and merger.  The predicted DTD can be approximated with a simple $t^{-1}$ power law, but there is significant variation between models and some metallicity dependence. The differences are due to a range of uncertain assumptions regarding all aspects of binary evolution: input parameter distributions, wind mass loss and stellar evolution, supernova remnant masses and natal kicks, the stability of mass transfer and the fraction of mass accreted by the companion, and the treatment of the common envelope phase \citep[e.g.,][]{Broekgaarden:2022}.

The aim of this paper is not to perform a parameter study of r-process enrichment using GCE models and assuming NSM rates and DTDs as in previous works \citep[e.g.,][]{cot19,mol21}.  Rather, we aim to derive an insight into binary evolution physics by considering what NSM yields and delay times would need to be in order to explain observed Milky Way r-process abundances with NSMs alone. \S 2 briefly describes our GCE models.  Model predictions and comparison to observational data are presented in \S 3. We discuss the implications for binary evolution models in \S 4. \S 5 gives our conclusions.

\section{Model}

We use the GCE code from \citet{kob00} but including updated nucleosynthesis yields of AGB stars and core-collapse supernovae from \citet[][hereafter K20]{kob20sr} and those of Type Ia supernovae (SNe Ia) from \citet{kob20ia}. The initial mass function (IMF) and the parameters for star formation histories are the same as in K20, which are determined to match other observational constraints in the solar neighborhood (Fig.\,1 of K20). In general these parameters are insensitive to elemental abundance tracks for constraining stellar physics (see Fig.\,A1 of \citealt{kob20ia}).

The core-collapse supernova yields are calculated based on stellar evolution and explosive nucleosynthesis by \citet{ume99}. For supernovae with explosion energies of $\sim 10^{51}$ erg, a mass cut determines the remnant mass, while a mixing-fallback model is adopted for hypernovae (with energies $>10^{51}$ erg), which mimics an aspherical explosion \citep{nom13}. These assumptions are chosen to match the resultant Fe yields with the observations of nearby supernovae (i.e., light curves and spectra; see \S 2.1.1 of K20), and the yield set can well reproduce the evolutionary trends of most elements.

\begin{table*}[t]
\begin{center}
\caption{Predicted ejecta masses and Eu yields for mergers with given component masses; all masses are in $M_\odot$}
\begin{tabular}{l|ccccccc|c}
\hline
NSMs & 1.2+1.8 & 1.35+1.8 & 1.45+1.45 & 1.6+1.6 & 1.75+1.75 & 1.45+2.91 & 1.4+5.08 & 1.3+1.3\\
\hline
\multicolumn{8}{c|}{$\alpha_{\rm vis}=0.02$} & (W14)\\
\hline
$M_{\rm ejecta}$  & 0.0276 &  0.0270  & 0.0370 &  0.00717 &  0.00813 & N/A    &  N/A & 0.01\\
$M({\rm Eu})$
 &  6.76E-05 &  6.14E-05  & 1.93E-04  & 2.80E-05  & 2.69E-05  & N/A  & N/A  & 1.64E-05\\
\hline
\multicolumn{8}{c|}{$\alpha_{\rm vis}=0.05$} & \\
\hline
$M_{\rm ejecta}$    & 0.0296 &  0.0290  & 0.0393  & 0.00747 &  0.00843 & 0.1021  &    0.1233 &\\
$M({\rm Eu})$ & 1.65E-04 &  1.58E-04 &  2.90E-04  & 3.93E-05  & 3.46E-05&   6.46E-04  & 9.41E-04 &\\
\hline
\multicolumn{8}{c|}{Selection of events from BPS} & \\
\hline
$M_{\rm tot}$ & $<3$ & $<3$ & $<3$ & $<3.4$ & $<4.7$ & (BH-NS) & (BH-NS)&\\
$q$ & $<0.70833$ & 0.70833-0.875 & $\ge0.875$ & - & - & - & - &\\
$M_1$ & - & - & - & - & - & $<4$ & $<6$ &\\
\hline
\end{tabular}
\tablecomments{W14 includes dynamical ejecta only, not winds after mergers.
The yields are associated with BPS events based on $M_{\rm  tot}$ and/or $q$ for NS-NS mergers, and based only on $M_1$ for BH-NS mergers. A dash means no constraints. For example, the 1.6+1.6 NS-NS model is applied for $3\le M_{\rm tot}/M_\odot<3.4$.
The maximum NS mass is set to be $2.0 M_\odot$ in the COMPAS model used here and $2.5 M_\odot$ for StarTrack.
}
\end{center}
\end{table*}

The average evolution of chemical composition in the interstellar medium (ISM) is calculated by integrating the contributions of all relevant enrichment sources as a function of metallicity and timescale (see formulae in \citealt{kob00}, or those of the single-star contribution in \citealt{mat21}). 
The timescales are determined simply from the lifetimes of stars depending on their mass (and metallicity) for AGB stars and core-collapse supernovae, while the DTDs of merging binaries depend on many factors and model assumptions.

In this paper DTDs are defined as the rates of NSMs and SNe Ia per unit time per unit stellar mass formed in a population of stars with a coeval chemical composition and age (simple stellar population, SSP): $R_{t}=d^2N/dt/dM_{*,{\rm init}}$, where $M_{*,{\rm init}}$ is
the initial mass of the SSP.

There are some observational constraints on the DTDs in present-day galaxies (with observations biased toward high metallicity) or the rates integrated over time and volume containing various types of stars/galaxies \citep[e.g.,][]{van96}. However, for GCE, we need DTDs or rates as a function of metallicity of the SSP. BPS models \citep[e.g.,][]{tut93,bel02,men14,kru18,vig18} can provide the necessary information for GCE, i.e., the rate of events per unit stellar mass formed as a function of delay time and metallicity. However, predictions vary among the BPS codes because of uncertain assumptions regarding poorly understood binary physics, such as mass transfer including common envelope evolution or supernova natal kicks. \citet{ded04} were the first to explore such binary effects in a GCE code.

\begin{figure*}\center
\includegraphics[width=15.5cm]{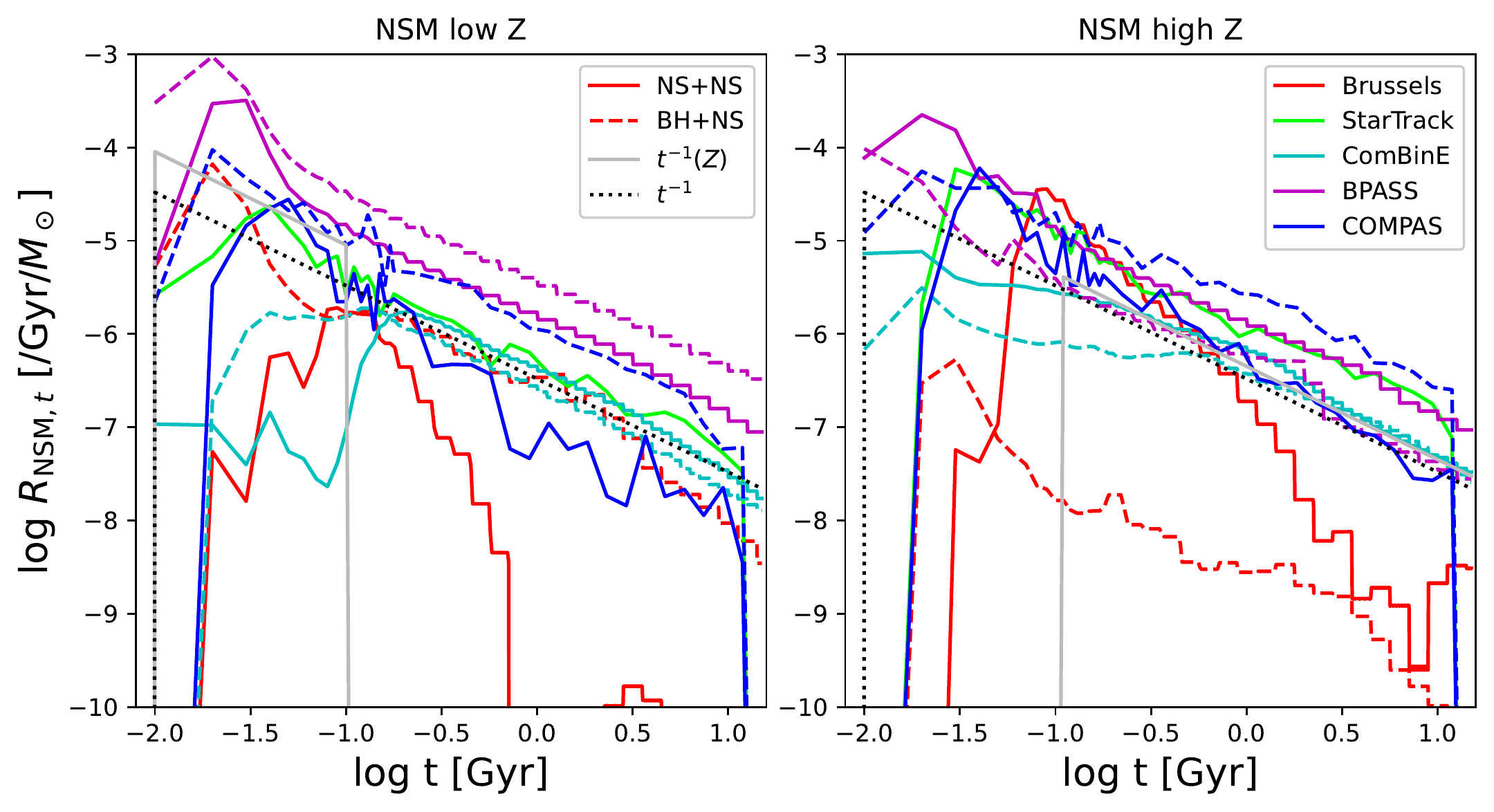}
\includegraphics[width=15.5cm]{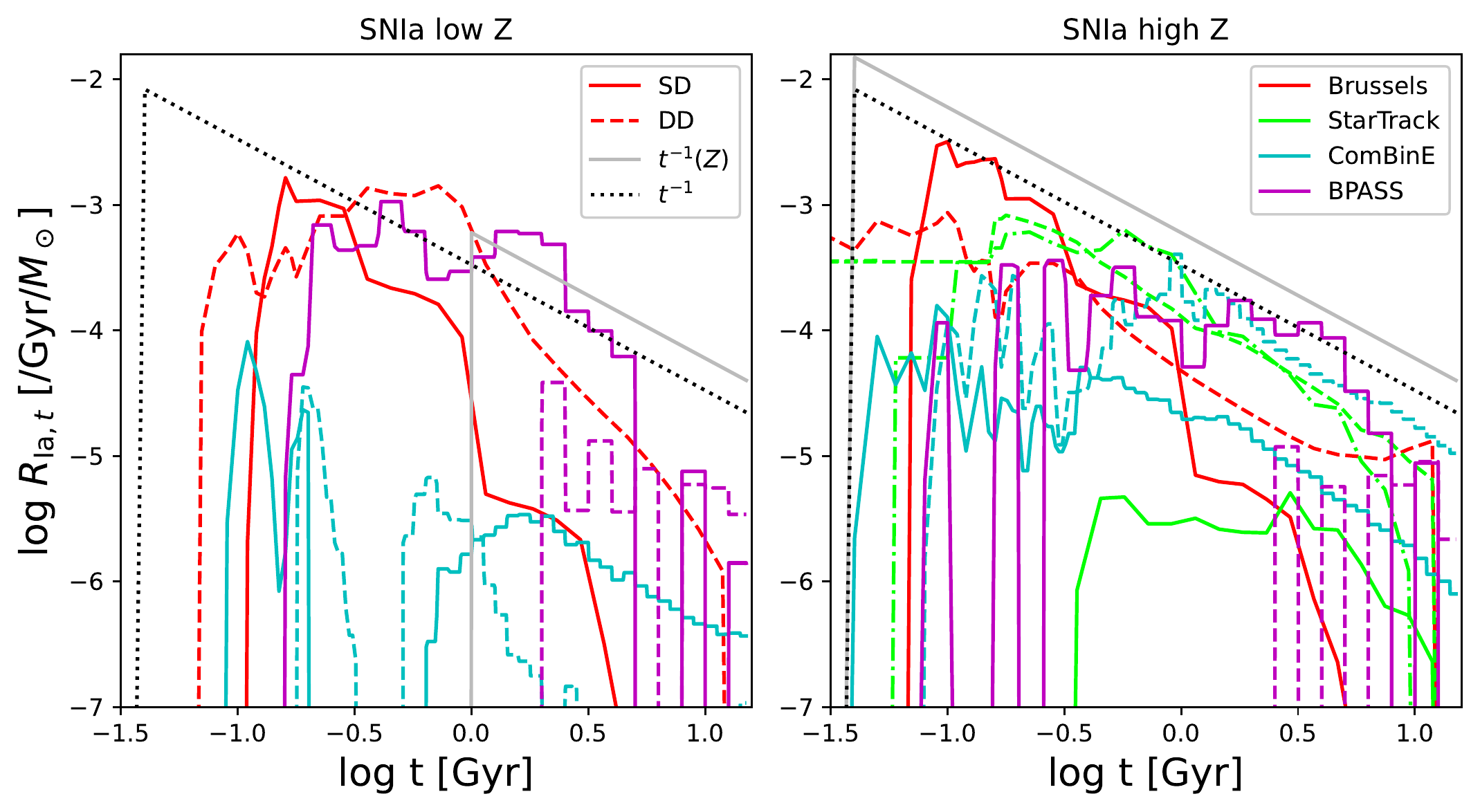}
\caption{\label{fig:dtd}
DTDs of NSMs and SNe Ia at low (left panel) and high (right panel) metallicities, 
calculated with various BPS codes.
Metallicity independent (black dotted lines) and dependent (gray solid lines, $Z=0.002$ and $0.02$) $t^{-1}$ power laws are also shown (see \S 3 for more details).
(Upper panels) The adopted BPS models are from:
\citet[red lines, $Z=0.002$ and $0.02$, Model 2]{men14,men16};
\citet[green lines, $Z=0.001$ and $0.02$, M30 model]{bel02,bel20};
\citet[cyan lines, $Z=0.0002$, $0.0088$]{kru18};
\citet{eld09}, \citet[magenta lines, $Z=0.002$ and $0.02$]{bri22};
and
\citet[blue lines, $Z=0.0014$, $0.014$]{man21}.
(Lower panels) The adopted BPS models with their preferred definitions of SNe Ia are from:
\citet[red lines, SD and DD with $M_{\rm tot}\ge1.4{\Msun}$, Model 22]{men10,men12};
\citet[green lines, SD (solid line), DD with $M_{\rm tot}\ge0.9{\Msun}$ (dashed line), and He accreted double detonations (dot-dashed line)]{rui09,rui14};
for ComBinE private communication based on \citet[cyan lines, DD only, with $M_{\rm tot}>1.36{\Msun}$ (solid lines) and $M_{\rm tot}\ge0.86{\Msun}$ (dashed lines)]{kru18} but with model `CE';
and \citet{eld09}, \citet[magenta lines, SD and DD]{bri22}.
}
\end{figure*}

Figure \ref{fig:dtd} shows the DTDs\footnote{Equivalently, the merger rates per unit stellar mass formed following a SSP.} of NSMs from various BPS codes, with the parameter sets recommended by the authors.  In K20, for NSMs, the DTDs of the standard model from \citet[][hereafter Brussels]{men14,men16} for $Z=0.002$ and $Z=0.02$ were used, assuming a 100\% binary fraction (which is also the case throughout this paper).
Because of the larger mass of progenitors, BH-NS mergers have shorter minimal delay times than NS-NS mergers, and thus they are potentially good candidates for the early r-process enrichment.
Moreover, at low metallicities, the rates of BH-NS mergers are predicted to be higher than those of NS-NS mergers, while at high metallicities predictions vary for different BPS simulations as follows.  
The BH-NS merger rate becomes lower than the NS-NS merger rate at high metallicities in Brussels, \citet[][hereafter ComBinE]{kru18}, and \citet[][hearafter BPASS]{bri22},
while the BH-NS rate is still slightly higher than the NS-NS merger rate in \citet[][hereafter COMPAS]{man21}.
For \citet[][hereafter StarTrack]{bel20}, the BH-NS rate is so low and sparse that it is not plotted in this figure.

As in K20, the nucleosynthesis yields are taken from \citet[][hereafter W14]{wan14}, and are calculated by post-processing a 3D-GR simulation of a NS-NS merger ($1.3M_\odot+1.3M_\odot$). Although the parameter dependence was discussed in K20 (see their \S 2.1.2), no yields were available for further investigation. In this paper, we also use nucleosynthesis yields from \citet{jus15}, which allow us to study the dependence of yields on NSM masses.
While W14 yields included only dynamical ejecta of the NSM, \citet{jus15} added the dominant BH torus ejecta driven by neutrinos and viscosity ($\alpha_{\rm vis}$)\footnote{The ejecta from a remnant hypermassive NS, created in systems with low-enough binary mass and/or stiff-enough nuclear EoS, are not included. Depending on its lifetime and yet poorly understood magnetohydrodynamic processes of angular momentum transport and energy dissipation, the hypermassive NS could give rise to the ejection of substantial amounts of matter, presumably with relatively low neutron-richness compared to the other ejecta components \citep{martin15,fujibayashi18}.}. 
The spin of the BH formed after NSMs and the mass of the torus are also given as parameters. While the dynamical ejecta consistently have low electron fraction $Y_{\rm e}$ and produce very heavy r-process elements, the secular (here called ``wind'') outflow component has relatively higher $Y_{\rm e}$ and has more impact on element production with $A\ltsim130$.
For the EoS of NSs, we use the models `SFHO' or `DD2' if available in \citet{jus15}, and if not, we re-scale those with `TM1' or `TMA' by using the dynamical ejecta masses for `SFHO' from \citet{bau13} and \citet{bau14}.

The Eu yields are summarised in Table 1, compared with the value (labelled W14) used in K20.
The ejecta mass is higher for systems with more extreme mass ratios, as shown in Fig.\,7 of \citet{bau13}.
For equal-mass mergers, the mass of dynamical ejecta is similar between W14 and \citet{jus15}; it is 0.00483 $M_\odot$ for the 1.35+1.35 $M_\odot$ merger model in \citet{bau13} and 0.0143 $M_\odot$ for the 1.45+1.45 $M_\odot$ model in \citet{jus15}, although the time when the mass is calculated is different; there is no clear separation between dynamical and wind components.
With the same EoS, \citet{rad18} found ejecta masses of $0.0035$ and $0.0004$ $M_\odot$ for 1.35+1.35 $M_\odot$ and 1.4+1.4 $M_\odot$ mergers, respectively\footnote{This difference arises because the two mass choices straddle the boundary for the prompt collapse of the merger product into a black hole for this EoS, leading to a bifurcation in the ejecta amounts and compositions.}. 
The main difference in the Eu yields is caused by the wind component added in \citet{jus15}, depending on the viscosity parameter $\alpha_{\rm vis}$; the torus mass was also assumed to be 0.1, 0.03, or $0.3M_\odot$ respectively for the first three models, the massive NS-NS mergers, and the NS-BH mergers in Table 1 (see \citealt{jus15} for the other parameters of the remnant models).
Although the results of  \citet{jus15} give $\sim$2-60 times larger Eu yields than W14, they also produce non-negligible amounts of Fe and O, and thus the [Eu/(O,Fe)] ratio is only larger by a factor of a few compared to W14.
Hence, it is important to include not only the Eu yields but also the yields of other elements in GCE.

When implementing these mass-dependent yields in GCE, we match BPS-modelled NSMs to the yields in Table 1 based on the total mass $M_{\rm tot}\equiv M_1+M_2$, the mass ratio $q\equiv M_2/M_1$, and the primary mass $M_1$ as described in the bottom section of the table. 
Whether or not BH-NS mergers eject significant mass and contribute to r-process nucleosynthesis depends on the BH mass, the mass and EoS of the NS, and on the spin of the BH and its tilt relative to the rotation axis of the binary \citep{bau14,Foucart:2018}.  If the BH mass is too high, NSs can directly plunge in without significant disruption; this was likely the case for both BH-NS mergers observed as gravitational-wave sources \citep{GW200105}.  We exclude BH-NS mergers with $M_1 \ge 6 M_\odot$ from contributing to GCE\footnote{These BH-NS mergers are included in Fig.\,\ref{fig:dtd}.};
this limit holds for compact NSs, but a high BH spin and a stiff EoS of the NS could permit ejecta for even more massive BHs.
The BH-NS models described in Table 1 assumed a BH dimensionless spin of $0.8$; with a smaller spin, the mass ejection becomes much smaller.
Following \citet{bau14}, an additional parameter $f_{\rm spin}$ is introduced to characterise the fraction of BH-NS mergers where the BH spin and tilt are favourable to sufficient matter ejection.
We apply the nucleosynthesis yields (of all elements) of the BH-NS models in Table 1 to a fraction $f_{\rm spin}$ of BH-NS mergers produced in BPS models in the relevant mass range.

When we use elemental abundances relative to Fe, the modelling of SNe Ia is also important. In K20, the progenitor model of \citet{kob09} is used, which is based on the theoretical calculation of optically thick winds from white dwarfs (WDs) by \citet{hac08}. The DTDs depend on the metallicity of the binary systems, and are shown in Fig.\,3 of \citet{kob09}.  These are similar to the ``observed'' DTDs from \citet{mao14} at high metallicities (see also Fig.\,12 of \citealt{kob20ia}). 
In this paper, we also use DTDs from BPS models for SNe Ia.
Figure \ref{fig:dtd} shows the DTDs for single degenerate (SD) and double degenerate (DD) systems from various BPS codes, again with the parameter sets recommended by the authors. Even for a given BPS code, the parameter sets are not necessarily the same as for NSMs (see Di Stefano et al. in prep. for details).
Note that \citet{hac08}'s WD wind effects are included only in Brussels.
The metal-dependent nucleosynthesis yields of near-Chandrasekhar-mass (near-Ch-mass) and sub-Ch-mass models from \citet{kob20ia} are adopted for the SD and DD (or DD+He) systems, respectively. For ComBinE, DTDs of DD systems are given as a function of the total mass, and 1.37, 1.2, 1.1, 1.0, and $0.9M_\odot$ WD yields are used respectively for $M_{\rm tot}>2.12$, 1.36, 1.16, 1.06, and $0.96M_\odot$. For the other BPS, since the mass dependence is not available, we add 1.2, 1.1, 1.0, and $0.9M_\odot$ WD yields, respectively, with 10\%, 40\%, 40\%, and 10\% contributions toward the total sub-Ch-mass SNe Ia. The results are not very sensitive to this averaging.

\section{Results}

Elemental abundances are being measured in the stellar atmospheres of nearby Milky Way stars for the past half century. Nowadays Galactic archaeology surveys with multi-object spectrographs are providing data for a million stars \citep[e.g.,][]{buder21}. However, in order to constrain stellar physics, elemental abundances with 3D non-local-thermodynamic-equilibrium (NLTE) analysis of higher resolution data are necessary \citep{kob20sr}.
Among r-process elements, Eu is the element accurately measured for the largest number of stars (see Fig.\,32 of K20).
Following \citet{hay19}, in order to remove the contribution from SNe Ia, we also show Eu abundances relative to $\alpha$ elements. Although Mg is the best observed $\alpha$ element, O is the element with the most robust nucleosynthesis yields (see \S 3.6 of K20) and is thus used in this paper.

Figure 2 shows the [O/Fe]–[Fe/H], [Eu/Fe]–[Fe/H], and [Eu/O]–[O/H] relations in the solar neighborhood.
The observational data are taken from K20 and \citet{yon21}, and the model predictions use the same W14 yields as in K20.
The black solid lines are the same as in Fig.\,37 of K20, which can well reproduce the average evolution in all panels, thanks to magneto-rotational supernovae\footnote{The B11$\beta$1.00 model from \citet{nis15} is adopted; see K20 for more details.}, which produce most of Eu in this model.
The other models consider NSMs as the only r-process site. 
Note that Brussels' DTDs for NSMs were already used in K20, but in this paper their DTDs of SD and DD SNe Ia are also used together with the metallicity-dependent yields of Ch-mass and sub-Ch mass SNe Ia from \citet{kob20ia}.
Similarly, for the other BPS models, DTDs of {\it both} NSMs and all available channels of SNe Ia are used, without applying any arbitrary shift.

\begin{figure}\center
\includegraphics[width=8.5cm]{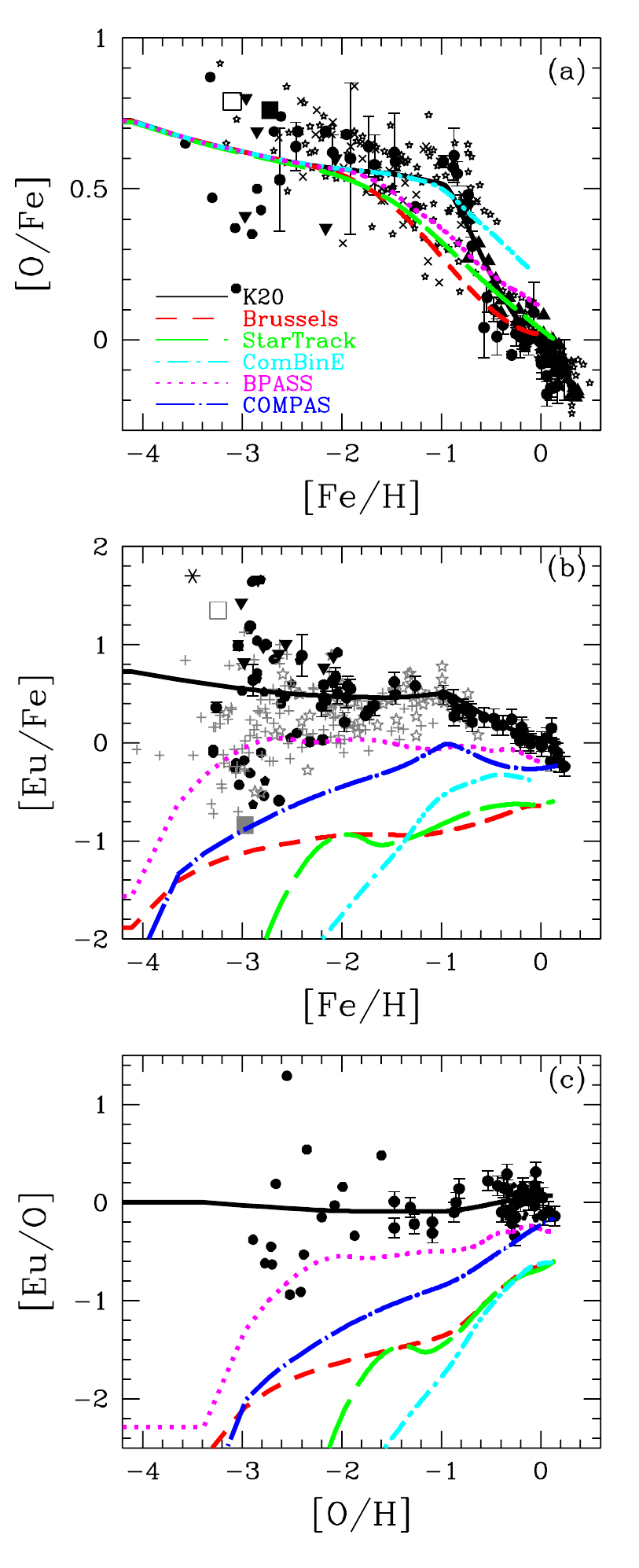}
\vspace*{-10mm}
\caption{\label{fig:eu1}
Evolution of elemental abundances in the solar neighborhood, calculated with our GCE code including DTDs in Fig.\,\ref{fig:dtd}.
The GCE code includes both SD and DD for SNe Ia, and both NS-NS and BH-NS for NSMs, unless noted below.
The BPS models are taken from:
\citet{men12}, \citet[red short-dashed lines]{men16};
\citet{rui14}, \citet[green long-dashed lines, also including He double det.]{bel20};
\citet[cyan dot-short-dashed lines, DD only]{kru18};
\citet[magenta dotted lines, NS-NS only]{bri22};
and \citet[blue dot-long-dashed lines]{man21}.
The nucleosynthesis yields are from W14, the same as in K20.
See K20 for the observational data sources.
The asterisk indicates the star found by \citet{yon21}.
}
\end{figure}

Figure \ref{fig:eu1}a shows that none of the BPS models can accurately reproduce the evolutionary change in [O/Fe] at [Fe/H] $\sim -1$.
Brussels, StarTrack, and BPASS models can reach the solar ratios (i.e., [O/Fe] $\sim 0$ at [Fe/H] $\sim 0$), but their predictions for typical SN Ia time delays are so short that [O/Fe] gradually decreases from [Fe/H] $\sim -2$. This is inconsistent with these observations, and also with the recent results from galactic surveys \citep[e.g.,][]{buder21}, which showed a clear `knee' at [Fe/H] $\sim -1$.
ComBinE (which includes DD only) gives a better match to the observations at [Fe/H] $\lesssim -1$, but the SN Ia rate is too low at higher metallicities. 
Note again that COMPAS does not provide a SN Ia DTD, so K20's SN Ia model is used, and thus the result is not plotted in the panel a.

The predicted [Eu/Fe] evolution in Figure \ref{fig:eu1}b is sensitive to the BPS models. BPASS shows the best result among the NSM-only models but the [Eu/Fe] ratio is $\sim 0.5$ dex lower than observed at [Fe/H] $\ltsim -1$ and the decrease in [Eu/Fe] from [Fe/H] $\sim -1$ to $\sim 0$ is not reproduced.
It is possible to fix the offset if the Eu/Fe ratio of the NSM yields is three times larger than in W14 (see \S 2).
In that case, however, the [Eu/Fe] ratio at [Fe/H] $\gtsim -1$ would be higher than observed.
COMPAS gives a lower [Eu/Fe] at [Fe/H] $\ltsim -1$ than BPASS, and ComBinE gives an even lower value, because the typical NSM delay timescales are too long. Brussels gives similar evolution as in BPASS but the [Eu/Fe] ratio is ten times lower than in BPASS, and the trend at [Fe/H] $\gtsim -1$ is not reproduced either. StarTrack predictions compare poorly with observations.
However, these results are affected by the over- or under- predictions of SNe Ia as shown in Fig.\,\ref{fig:eu1}a.

Meanwhile, [Eu/O] evolution is shown in Figure \ref{fig:eu1}c. The number of observational data points is reduced but the two best data sets remain, namely, the NLTE analysis from \citet{zhao16} at [O/H] $\gtsim -1.5$ and LTE analysis from \citet{spi05} with 3D correction of $-0.23$ dex at [O/H] $\ltsim -1.5$.
Although BPASS models are the closest to the observations, the systematic $\sim 0.5$ dex offset is still seen, which can be solved with a higher Eu yield. However, [Eu/O] ratio shows an increase from [O/H] $\gtrsim -1$, which would give a too high [Eu/O] ratio.  ComBinE gives the worst result when the SN Ia contribution is removed by plotting the [Eu/O] ratio.

Note that with a higher Eu yield per event, [Eu/(O,Fe)] ratios can be systematically higher in Figure \ref{fig:eu1}. This will help for some of the BPS models to match the solar ratios (at [Fe/H] $=0$) but none of these BPS models can reproduce the observed evolutionary tracks; [Eu/(O,Fe)] should be increased more at early times than at present, if NSMs are the major r-process site.

\begin{figure}\center
\includegraphics[width=8.5cm]{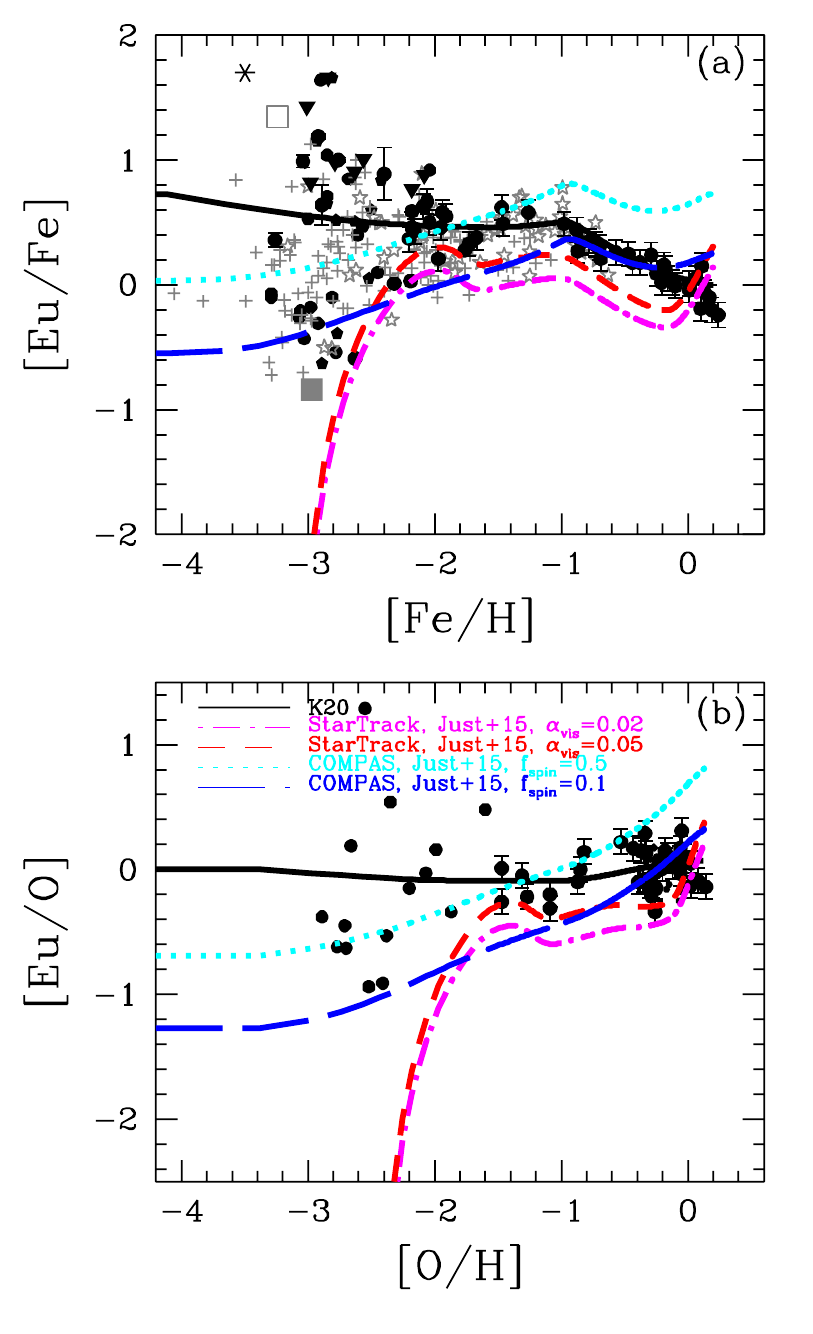}
\caption{\label{fig:eu3}
The same as Fig.\,\ref{fig:eu1} but with mass-dependent yields from \citet{jus15}, using BPS of NSMs from \citet[magenta and red lines]{bel20} and \citet[cyan and blue lines]{man21}.
The SN Ia model is the same as in K20.
See the main text for additional parameters.
}
\end{figure}

In Figure \ref{fig:eu3}, we include the BPS-predicted mass distribution of NSMs together with the mass-dependent yields (of all elements) from \citet{jus15}. We need to know the delay times (time from star formation to coalescence) and masses of the compact objects; we have such necessary information for GCE only from COMPAS\footnote{We use 2 models with $Z=0.0014$ and 0.014 in \citet{man21} and exclude events with `Optimistic\_CE'=True or `Immediate\_RLOF$>$CE'=True. The COMPAS output already assumed a maximum neutron star mass of $M_{\rm NS,max}=2.0M_\odot$.} and StarTrack\footnote{We use the M30 model in \citet{bel20} but with $Z=0.0001$, 0.0005, 0.001, 0.004, 0.008, 0.01, 0.02, and 0.03 and the events with `hece'=0, which yields a NS-NS merger rate consistent with gravitational-wave observations. $M_{\rm NS,max}=2.5M_\odot$ is applied in the GCE code.}.
The results depend on additional parameters: the Shakura-Sunyaev viscosity $\alpha_\mathrm{vis}$ of the BH torus plasma in \citet{jus15} and fraction of BH-NS mergers that have spin and tilt suitable for sufficient mass ejection after mergers, $f_{\rm spin}$, in \citet{bau14}. SNe Ia are also included with the same model as in K20.

COMPAS has a high BH-NS merger rate and the nucleosynthesis predictions are thus highly sensitive to the $f_{\rm spin}$ parameter.
However, this model yields either too little Eu at low metallicities (with $f_{\rm spin}=0.1$, blue lines) or too much Eu at high metallicities (with $f_{\rm spin}=0.5$, cyan lines). $\alpha_{\rm vis}=0.02$ is assumed for NS-NS mergers in these two models.

StarTrack has a much lower BH-NS merger rate, and thus the result is insensitive to the spin+tilt parameter ($f_{\rm spin}=0.5$ is assumed in these two models) but depends on the viscosity parameter.
Higher $\alpha_{\rm vis}$ for NS-NS mergers (red lines) results in higher Eu abundances but these are still not large enough to match the observational data.
The sharp increase in Eu for super-solar metallicities is also problematic.
Note that the value of $f_{\rm spin}$ estimated in BPS studies is much smaller \citep{Drozda:2020}.

\begin{figure}\center
\includegraphics[width=8.5cm]{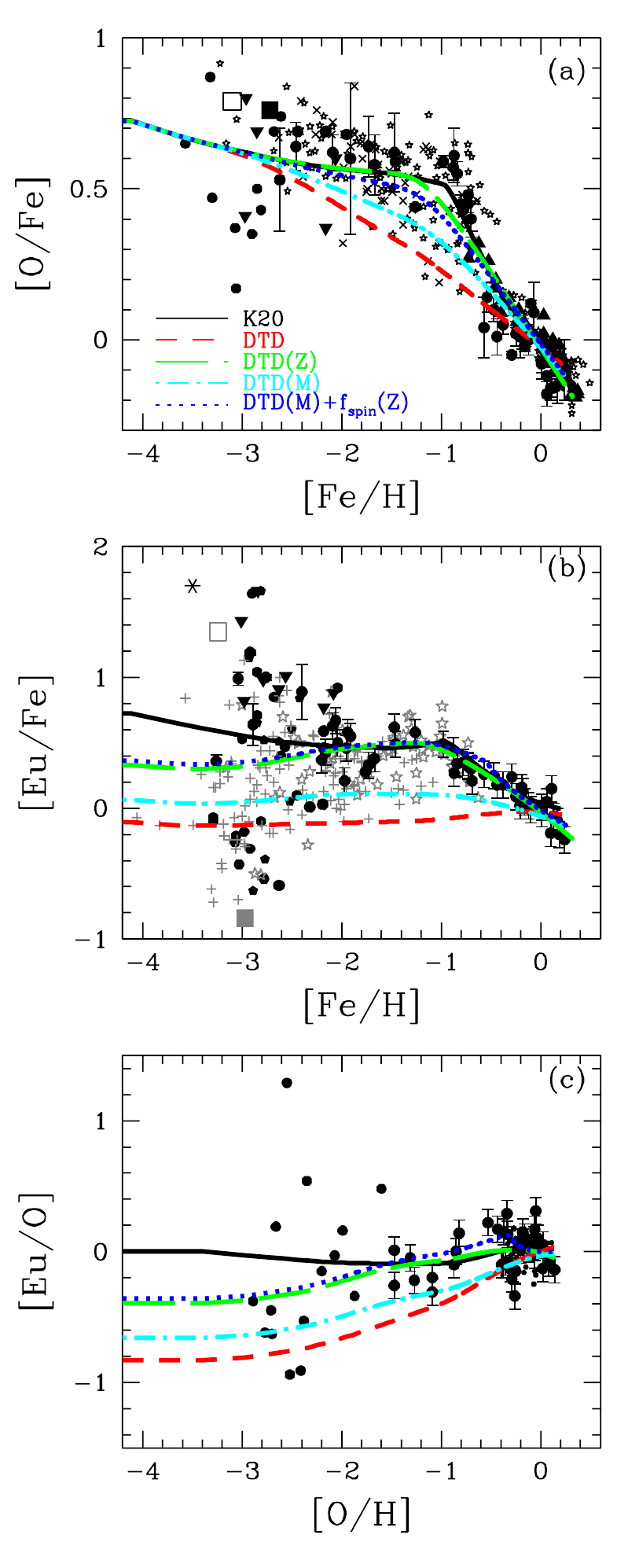}
\caption{\label{fig:eu2}
The same as Fig.\,\ref{fig:eu1} but with a single $t^{-1}$ power law (red short-dashed lines), metallicity-dependent (green long-dashed lines), and mass-dependent (cyan dot-dashed lines) $t^{-1}$ power laws. The blue dotted lines also assume a metallicity-dependent $f_{\rm spin}$ in addition to the mass-dependent DTDs. See the main text for the assumed DTDs and for the selected sub-set of SNe Ia and NSMs yields from \citet{kob20ia} and \citet{jus15}, respectively.
}
\end{figure}

Having seen that existing BPS models struggle to reproduce the observations, we can return to our original question: If NSMs were the unique r-process site, what kind of DTDs could reproduce the observational data? To answer this question, we assume $t^{-1}$ power-laws with fixed ranges for the DTD. The range and the height (i.e., rate) are arbitrarily determined in order to match the observations in Figure \ref{fig:eu2}.
Predictions for the following single power-law are shown with the red short-dashed lines:
\begin{eqnarray}
\label{eq:DTD}
\begin{array}{ccl}
R_{{\rm NSM},t} &= 3.3\times10^{-7}\,t^{-1}\,M_\odot^{-1} & \mbox{for 0.01-20 Gyr}\\
R_{{\rm Ia},t} &= 3.33\times10^{-4}\,t^{-1}\,M_\odot^{-1} & \mbox{for 0.04-20 Gyr} .
\end{array}
\end{eqnarray}
Here the 1.45+1.45 $M_\odot$ yields are used, and the sub-Ch mass fraction to total SNe Ia is set to be 25\%; these assumptions changes the normalization only.
These single power-law functions cannot reproduce observations.
The single power-law function gives a reasonable approximation for NSMs and SNe Ia from DD, but perhaps not for the other SNe Ia \citep[][for more discussion]{rui11}.
The mismatches compared with the observations, apparent in Figure \ref{fig:eu2}, are mainly caused by the minimum delay-times for both SNe Ia and NSMs; SNe Ia occur too early to keep the [$\alpha$/Fe] plateau, and NSMs occur too late to create the [Eu/(O,Fe)] plateau. 
A flatter power than $-1$ gives a slightly shallower slope in the panels a and c, but it does not solve these problems.

However, it is possible to reproduce the data if the DTDs strongly depend on metallicity (green long-dashed lines in Figure \ref{fig:eu2}) as indicated by the gray solid lines in Figure \ref{fig:dtd}:
\begin{eqnarray}
\label{eq:zDTD}
\begin{array}{ccl}
R_{{\rm NSM},t}(Z=0.002)&=9\times10^{-7}\,t^{-1}\,M_\odot^{-1} & \mbox{for 0.01-0.1 Gyr}\\
R_{{\rm NSM},t}(Z=0.02)&=4.5\times10^{-7}\,t^{-1}\,M_\odot^{-1} & \mbox{for 0.1-20 Gyr}\\
R_{{\rm Ia},t}(Z=0.002)&=6\times10^{-4}\,t^{-1}\,M_\odot^{-1} & \mbox{for 1-20 Gyr}\\
R_{{\rm Ia},t}(Z=0.02)&=6\times10^{-4}\,t^{-1}\,M_\odot^{-1} & \mbox{for 0.04-20 Gyr} .
\end{array}
\end{eqnarray}
As for Eq.\ref{eq:DTD}, the 1.45+1.45 $M_\odot$ yields and the 25\% sub-Ch mass fraction adopted.
These functions given here are interpolated linearly in $\log Z$ at intermediate metallicities, while constant values are extrapolated at metallicities higher (lower) than $Z=0.02$ (0.002).

For SNe Ia, optically thick winds from WDs can explain such a metallicity dependence of the rate \citep{kob98}, although this effect is not fully included in existing BPS models.
Metallicity dependent yields of Ch and sub-Ch mass SNe Ia  \citep{kob20ia} are also used throughout this paper.

For NSMs, nucleosynthesis yields at given NS/BH masses (with a given spin) are not expected to depend on the metallicity.
The NSM DTD may strongly depend on the masses of compact objects, and one may think that the mass dependence could give a similar effect to the metallicity dependence.
However, we find that the mass dependence alone (cyan dot-dashed lines) cannot reproduce the data in Figure \ref{fig:eu2}.
To show this, we assume an extreme case of DTD that strongly depends on the mass, using the 1.6+1.6 $M_\odot$ and 1.4+5.08 $M_\odot$ yields with $f_{\rm spin}=0.1$, respectively, for NS-NS and BH-NS mergers, and $1.37M_\odot$ and $1.0M_\odot$ yields, respectively, for Ch and sub-Ch mass SNe Ia, with 25\% sub-Ch mass fraction:
\begin{eqnarray}
\label{eq:mDTD}
\begin{array}{ccl}
R_{{\rm NSM},t}({\rm BH\textrm{-}NS})&=1\times10^{-6}\,t^{-1}\,M_\odot^{-1} & \mbox{for 0.01-1 Gyr}\\
R_{{\rm NSM},t}({\rm NS\textrm{-}NS})&=1\times10^{-6}\,t^{-1}\,M_\odot^{-1} & \mbox{for 0.1-20 Gyr}\\
R_{{\rm Ia},t}({\rm Ch})&=6\times10^{-4}\,t^{-1}\,M_\odot^{-1} & \mbox{for 1-20 Gyr}\\
R_{{\rm Ia},t}({\rm subCh})&=2\times10^{-4}\,t^{-1}\,M_\odot^{-1} & \mbox{for 0.04-20 Gyr} .
\end{array}
\end{eqnarray}
Since BH-NS mergers yield more Eu than NS-NS mergers in Table 1, this model gives a higher [Eu/Fe] ratio at early times than the single power-law model. However, at later times, [Eu/O] keeps increasing and [Eu/Fe] shows a shallower decrease than observed. This is because star formation is ongoing, BH-NS mergers keep happening, and NS-NS mergers also start contributing at higher metallicities. Any changes of the range and normalization of the DTD do not help.
It is not possible to reproduce the [O/Fe] ratios either; since our averaged Fe yield\footnote{The mass-dependent yields are averaged with the WD mass distribution in \S 2.} from sub-Ch-mass SNe Ia is smaller than that of Ch-mass SNe Ia, the [O/Fe] ratio is higher than the single power-law model at [Fe/H]$<0$, but is still lower than observed.

If BH spins strongly depend on metallicity, it becomes possible to reproduce the data. The blue dotted lines are for the mass-dependent DTDs in Eq.\ref{eq:mDTD}, but assuming a higher $f_{\rm spin}$ at low metallicities as follows:
\begin{eqnarray}
\begin{array}{ccl}
f_{\rm spin} & = 0.1 & \mbox{ for } Z>0.006\\
& = 0.2  & \mbox{ for } Z\le0.006.
\end{array}
\end{eqnarray}
Here the fraction of sub-Ch mass SNe Ia of total SNe Ia is assumed to be only 5\%; the coefficients are (9.5, 0.5), instead of (6, 2) in Eq.\ref{eq:mDTD}.
Note that the sub-Ch mass fraction changes the normalization only for the models with Eq.\ref{eq:DTD}, while it affects the shape of model curves with Eqs. \ref{eq:zDTD} and \ref{eq:mDTD}.
The sub-Ch mass fraction is very important for other elemental abundances, namely Mn, and found to be less than 25\% from the Mn constraint in \citet{kob20ia}.

\section{Discussion}

The GCE models we use in this paper to explore the parameter space of NSMs can well predict the average evolution of elemental abundances but not the scatter. It is known that r-process elements show a large scatter particularly at low metallicities, which is caused by the rareness of the enrichment sources \cite[e.g.,][]{arg04}.
There is also a selection bias of the observational data at [Fe/H] $\ltsim -3$, where
a few stars show very high r-process abundances, called r-II stars; the most metal-poor one (asterisk in Fig.~\ref{fig:eu1}) is the star in \citet{yon21} at [Fe/H] $=-3.5$. Although these stars are well studied to obtain abundance patterns, the number of such stars is small (see Fig.14 of \citealt{yon21b}).
Theoretically, the scatter can be predicted with stochastic GCE models \citep{ces15,weh15}, or more sophisticated, chemodynamical simulations of galaxies \citep{hay19,van20,van22}\footnote{\citet{van22} included the effect of natal kicks in their hydrodynamical simulations. With an `optimized' NSM DTD, the average r-process abundance can be as high as observed, but with kicks the scatter becomes too large.}.
Using the same yield set as in K20, \citet{hay19} showed that the model with magneto-rotational supernovae reproduces the observed scatter of [Eu/(O,Fe)] well, while the model with NSMs shows a too large scatter with low [Eu/(O,Fe)] ratios.
This is because the timescale of magneto-rotational supernovae is similar to that of Fe (and O)-producing core-collapse supernovae, while the NSMs are not associated with significant Fe (and O) production.
Therefore when we compare the GCE models to the observed elemental abundances in Figures \ref{fig:eu1}-\ref{fig:eu2}, it should be noted that the K20 model would scatter around the line, while the NSM-only models tend to scatter below the line.

Our analysis indicates that existing BPS models of NSMs struggle to reproduce the observed Eu abundances within our adopted yield models (Figs.\,\ref{fig:eu1} and \ref{fig:eu3}). The metallicity-dependent DTDs required to match observations are very different from those in existing BPS models (Figs.\,\ref{fig:dtd} and \ref{fig:eu2}).  Perhaps this should not come as a great surprise given that most BPS models do not accurately recover the Galactic double neutron star population properties.  For example, the simulated mass distributions are often inconsistent with observed ones, which is likely a statement about imperfect recipes for translating the progenitor properties to remnant masses after supernovae \citep{Fryer:2012,vig18, man21}.  Meanwhile, Eu yields appear to depend very sensitively but uncertainly on the masses of merging objects, as shown in Table 1 and associated discussion, as well as on other imperfectly constrained quantities such as the electron fraction \citep{kullmann22a}, or on nuclear physics input such as fission fragment distribution and nuclear mass model \citep[e.g.,][]{Eichler:2015, Holmbeck:2019,kullmann22b}.  Together, the likely flaws in population synthesis mass models and the uncertainty in NSM-mass-dependent yields make these comparisons challenging.

BPS models also struggle to reproduce the period-eccentricity distribution of Galactic double neutron stars, particularly the bifurcation in eccentricities for short orbital period systems \citep{AndrewsMandel:2019}.  This is, perhaps, more concerning than the mass distribution mismatch.  Firstly, since the period and eccentricity of a double neutron star system after formation is directly linked to the delay time until merger through gravitational-wave emission \citep{Peters:1964}, this means that predicted DTDs are also likely to be inaccurate. Secondly and more fundamentally, it is an indication of significant uncertainties in the late stages of evolution preceding the formation of the second neutron star, particularly case BB mass transfer from a stripped expanding post He main sequence secondary onto the NS primary and the subsequent ultra-stripped supernova \citep{Tauris:2015}, as well as
an uncertainty in the natal kicks of NSs and low-mass BHs \citep[e.g.,][]{Igoshev:2021,richards22,Kapil:2022}.

In particular, the amount of mass stripped during case BB mass transfer, the dynamical stability of the mass transfer (whether or not the binary experiences common envelope evolution) and thus the ultimate binary separation, the remnant masses and the delay time until merger may well prove to depend on metallicity.  The discovery of GW190425 \citep{abb20a}, a NSM with a total mass significantly higher than any Galactic double neutron stars, hints at the possibility that there may exist a population of more massive NSMs with very tight orbits and short delay times and/or at lower metallicities, thus evading observation as Galactic radio pulsars but detectable as gravitational-wave sources (\citealt{Safarzadeh:2020,Galaudage:2021DNS,VignaGomez:2021DNS}; see also \citealt{kruckow20}).  For example, if lower-metallicity environments yield more massive stripped He stars after the first common-envelope phase, the subsequent case BB mass transfer would occur in a system with a more extreme mass ratio, leading to more binary hardening and a shorter delay time.
Reduced mass loss in lower-metallicity environments could also produce tighter binaries and enhanced tidal spin-up, leading to more rapidly spinning BHs \citep[e.g.,][]{Bavera:2020} in NS-BH mergers, supporting our metallicity-dependent $f_{\rm spin}$ variation.  However, many BPS models suggest that the BH in merging NS-BH binaries typically almost always forms when the binary is still wide \citep[e.g.,][]{Broekgaarden:2021,Debatri:2022}, reducing the opportunity for tidal spin-up of its progenitor.

This motivates the ad-hoc models proposed at the end of the previous section.  The full space of allowed options is larger than we have considered: in addition to the freedom of choosing both mass- and metallicity-dependent DTDs, one could also vary the shape of the DTD (we did not explore this and restricted ourselves to $t^{-1}$ power laws; see also \citealt{dvorkin21}).  As mentioned above, more complex correlations, such as the possibility that low-metallicity environments yield both shorter DTDs and higher masses (and thus, perhaps, more Eu per NSM) are plausible.  However, at least some of the models, such as the metallicity-dependent DTD model, are largely consistent with Eu abundance observations while retaining plausible parameters.   For example, the total number of NSMs integrated over the age of the Universe is roughly 15-20 per million solar masses of star formation in both the single power law and metallicity-dependent DTD models.  This is consistent with $\sim 30$ NSMs per million years in the Galaxy, a value that matches inferred rates from Galactic radio pulsars \citep[e.g.,][]{Pol:2020}, and with around 300 NSMs per Gpc$^3$ per year, which falls within the range of inferred rates from gravitational-wave observations \citep{GWTC3:pop}.  

The contribution of BH-NS binaries to Eu yields is even less certain.  In order to contribute significant r-process nucleosynthesis, such mergers must disrupt the neutron star before it plunges through the innermost stable circular orbit (ISCO) of the BH, otherwise, insufficient material is left outside the BH.  The radius of the ISCO scales linearly with the BH mass for a non-spinning (Schwarzschild) BH, but the tidal disruption radius scales only with the $1/3$ power of the BH mass for sufficiently massive BHs.  Therefore, if the BH is too massive, the neutron star will generally plunge in before it is disrupted.  Disruption is easier if the BH is rapidly spinning and the inspiral is prograde, as the ISCO moves in (the ISCO radius can be up to a factor of 6 smaller for a maximally spinning Kerr BH than a Schwarzschild one).  The condition on mass retention and nucleosynthesis is thus a function of the BH mass, spin and orbital tilt angle, and neutron star mass and EoS \citep{bau14,Foucart:2018}. Given the uncertain mass and spin distribution of BHs in merging BH-NS systems \citep[e.g.,][]{Broekgaarden:2021}, it is unclear whether a significant fraction of such mergers could contribute to r-process nucleosynthesis. Thus, while some short gamma-ray bursts and perhaps kilonovae have been conjectured to be associated with BH-NS mergers \citep{Troja:2008,Berger:2014,Li:2017,Gompertz:2020}, other authors have argued that the contribution of BH-NS mergers to short gamma-ray bursts is likely low \citep[e.g.,][]{Drozda:2020}.  For example, it is probable that neither of the gravitational-wave observed BH-NS mergers would have experienced a kilonova \citep{GW200105}.

Unfortunately, other observations are not yet sufficient to strongly constrain the merger masses and delay times of NSMs.  Only a handful of mergers involving neutron stars have been so far observed with gravitational waves, making population inference challenging; furthermore, even inference on individual events may be sensitive to model assumptions \citep[e.g.,][]{MandelSmith:2021}.  Kilonova observations not connected with gravitational-wave sources \citep[e.g.,][]{Tanvir:2013} could ultimately provide further constraints on both the NSM rates (see \citealt{MandelBroekgaarden:2021} and references therein) and ejecta composition with more events, as expected in the Vera Rubin Observatory LSST data \citep{Andreoni:2019}.  Several attempts have been made to constrain DTDs using short gamma-ray bursts \citep[e.g.,][]{Virgili:2011,WandermanPiran:2015}.  However, such inference is sensitive to selection effects on both the prompt burst, X-ray afterglow and the later optical afterglow detection necessary for host galaxy identification.  Short gamma-ray bursts can also be used to constrain the mass distribution of merging objects since a jet launch sets requirements on the masses, but these studies are not yet conclusive \citep{Sarin:2022,Salafia:2022}.

The presence of multiple uncertainties have led to explorations of the r-process abundances as constraints on the mass ratio distribution of NSMs \citep{bau13,Holmbeck:2021} or even on the NS EoS \citep{rad18,Holmbeck:2022}.  Placing stronger constraints will also require more detailed and faithful models of nucleosynthesis yields from mergers as a function of the masses of the merging objects.  
There are potentially sharp features in the ejecta masses and associated r-process abundances.  The location of these features is not precisely known due to a combination of uncertainties in the merger models and the unknown EoS.  Combined with the imprecise mass distribution predictions from BPS models, this creates an extra challenge for the kind of analysis undertaken here.  Moreover, our notion of a ``match'' to observations that suffer from significant scatter is a somewhat imprecise visual averaging.  This is partially due to the challenging observational selection effects when measuring abundances and partly to choices made in which stars to follow up.
A homogeneous, non-biased observation of neutron capture elements for a wide range of metallicity (with NLTE analysis) would facilitate more robust comparisons between models and observations.

\section{Conclusions}

We investigate whether NSMs alone can explain the r-process enrichment of the Milky Way. In particular, we explore whether NS-NS and especially NS-BH mergers in the early Universe can rapidly produce r-process elements instead of magneto-rotational supernovae.
Provided that NSMs are the leading site of r-process nucleosynthesis, we constrain NSMs by comparing GCE models to the observed elemental abundances in the solar neighborhood, namely, [O/Fe]--[Fe/H], [Eu/Fe]--[Fe/H] and [Eu/O]--[O/H] relations.
As clearly indicated by the decreasing trend of [Eu/Fe] toward higher metallicities, DTDs of NSMs and SNe Ia should be very different.
We show that none of the BPS used in this paper, i.e., COMPAS, StarTrack, Brussels, ComBinE, and BPASS, can reproduce the observations (Fig.\,\ref{fig:eu1}); the DTDs of NSMs predicted from BPS have too long delay times, while those of SNe Ia have too short delay times.

We propose metallicity-dependent DTDs of SNe Ia and NSMs with $t^{-1}$ power-laws that can explain the observations (Fig.\,\ref{fig:eu2}).
BH-NS mergers could play an essential role in the early chemical enrichment in the Milky Way if they dominate at low metallicities and are rare at high metallicities, and if a sufficient fraction of BH-NS mergers are able to produce significant mass ejections and contribute to r-process nucleosynthesis.  However, the necessary BH-NS DTDs are also inconsistent with existing BPS models (Fig.\,\ref{fig:eu3}).

\begin{acknowledgments}
CK thanks E.~De Donder for discussions at a conference in Vulcano in 1999, which inspired some parts of this paper.
CK acknowledges funding from the UK Science and Technology Facility Council (STFC) through grant ST/R000905/1 \&  ST/V000632/1.
CK also acknowledges the Australian Research Council Centre of Excellence for All Sky Astrophysics in 3 Dimensions (ASTRO 3D), through project number CE170100013.
The work was also funded by a Leverhulme Trust Research Project Grant on ``Birth of Elements''.
IM acknowledges support from the Australian Research Council Centre of Excellence for Gravitational Wave Discovery (OzGrav), through project number CE17010004. IM is a recipient of the Australian Research Council Future Fellowship FT190100574.
KB acknowledges support from the Polish National Science Center (NCN) grant Maestro (2018/30/A/ST9/00050).
SG is FRS-F.N.R.S. research associate.
HTJ acknowledges support by the Deutsche Forschungsgemeinschaft (DFG, German Research Foundation) through Sonderforschungsbereich (Collaborative Research Center) SFB-1258 ``Neutrinos and Dark Matter in Astro- and Particle Physics (NDM)'' and under Germany's Excellence Strategy through Cluster of Excellence ORIGINS (EXC-2094)-390783311.
OJ acknowledges support by the European Research Council (ERC) under the European Union's Horizon 2020 research and innovation program under grant agreement No. 759253 and by the Deutsche Forschungsgemeinschaft (DFG, German Research Foundation) - Project-ID 279384907 - SFB 1245.
AJR acknowledges financial support from the Australian Research Council under award number FT170100243. 
\end{acknowledgments}

\bibliography{ms}{}
\bibliographystyle{aasjournal}

\end{document}